\documentstyle[12pt]{article}

\newcommand{\textlineskip}{\baselineskip=18pt}

\newcommand{\tcaption}[1]{                      
        \addtocounter{table}{1}
         {{\tenrm\offinterlineskip Table~\thetable . #1} }\hfil\break }

\newcommand{\fcaption}[1]{
        \addtocounter{figure}{1}
         {{\tenrm Fig.~\thefigure . #1} }\hfil\break }

 1
 1
 1

\def\qed{\hbox{${\vcenter{\vbox{                          
   \hrule height 0.4pt\hbox{\vrule width 0.4pt height 6pt
   \kern5pt\vrule width 0.4pt}\hrule height 0.4pt}}}$}}

\newcommand{\be}{\begin{eqnarray}}
\newcommand{\ee}{\end{eqnarray}}
\newcommand{\dslash}{\partial \hskip -0.5em /}
\newcommand{\Dslash}{D \hskip -0.7em /}

\newcommand{\tr}{{\rm tr}}
\newcommand{\Tr}{{\rm Tr}}

\newcommand{\La}{{\cal L}}
\newcommand{\A}{{\cal A}}

\newcommand{\T}{{\cal T}}
\newcommand{\ie}{{\it i.e.}\ }
\newcommand{\eg}{{\it e.g.}\ }

\input psfig

\begin{document}
\normalsize\textlineskip
\baselineskip=22pt
{\thispagestyle{empty}}
\setcounter{page}{1}

\rightline{UNITU-THEP-16/1993}
\rightline{December 1993}
\rightline{hep-ph/9312269}
\vspace{1cm}
\centerline{\Large \bf Strange S--Wave Excitation of the
NJL Soliton$^\dagger $}
\vspace{0.5cm}
\centerline{H.\ Weigel$^\ddagger $, R.\ Alkofer and H.\ Reinhardt}
\vspace{0.2cm}
\centerline{Institute for Theoretical Physics}
\vspace{0.2cm}
\centerline{T\"ubingen University}
\vspace{0.2cm}
\centerline{Auf der Morgenstelle 14}
\vspace{0.2cm}
\centerline{D-72076 T\"ubingen, FR Germany}

\vspace{2cm}
\centerline{\bf Abstract}
\vspace{0.5cm}

The recently developed method to investigate mesonic fluctuations
off the chiral soliton in the Nambu--Jona--Lasinio model is
applied to kaons in the S--wave channel. It is shown that for
commonly accepted choices of parameters the presence of the soliton
and the lack of confinement in this model provide an unphysical
(valence) quark threshold which is lower than the kaon mass.
Choosing parameters such that this threshold is avoided a bound
state is obtained in the S--wave channel. After semiclassical
quantization this state acquires the quantum numbers of an odd
parity $\Lambda$ and is predicted at about 420MeV above the
nucleon mass.
\noindent

\vfill
\noindent
$^\dagger $
{\footnotesize{Supported by the Deutsche Forschungsgemeinschaft (DFG)
under contract  Re 856/2-1.}}
\newline
$^\ddagger $
{\footnotesize{Supported by a Habilitanden--scholarship of the DFG}}
\eject

\normalsize\textlineskip
\section{Introduction}

\medskip
The quark structure of baryons is still poorly understood. Nowadays,
two distinct or complementary pictures of the nucleon exist, namely,
either as a bound state of three quarks or as a solitonic lump of
meson fields, as \eg the Skyrmion. Whereas the first picture appeals
to everybody's intuition the latter seems more appropriate because
it provides quite a natural description of what has become famous as
the ``proton spin crisis"\cite{br88}. On the other hand we know that quark
degrees of freedom are important for baryon observables. In this
respect, it is quite encouraging that a simple quark model like that
of Nambu and Jona--Lasinio (NJL)\cite{na61} allows for solitonic
solutions\cite{re88}. Recent investigations within this model provide
some support for the soliton picture of baryons\cite{al92}.

An interesting intermediate position between baryons like nucleon and
$\Delta$ which are dominantly composed out of light quarks (up, down)
and baryons containing heavy quarks (charm, bottom, top?) is taken by
the hyperons which have a large strange component. To understand their
substructure one might assume the idea of considering \eg the $\Lambda$
as a bound state of the nucleon and an anti--kaon. Although this idea
is quite old\cite{go56} it has been revived in a certain sense by
Callan and Klebanov who described hyperons as kaonic bound states in
the Skyrmion background\cite{ca85}. Despite the phenomenological success
of this idea this description is somewhat unsatisfactory because
information about the strange quark in the hyperon can only be extracted
very indirectly. Recently, we have been able to describe the low--lying
normal parity hyperons as kaonic P--wave bound states in the NJL--soliton
background\cite{we93a,we93b}. This model provides direct access to the
strange quark(s) in the hyperon, this fact not only concerns the
valence level but also the Dirac sea.

To learn about the quark substructure of hyperons it may be even more
profitable to consider the ``wrong" parity hyperons, \eg the $\Lambda$
(1405) resonance. This has been done for the Skyrmion by investigating
the kaonic S--wave bound state\cite{da89}. In this paper we will examine
the corresponding bound state for the NJL soliton emphasizing aspects
related to the strange valence quark and/or Dirac sea. In section II
we derive the Bethe--Salpeter equation for the S--wave bound state and
the coupling strength of this mode to the collective (iso-) rotations.
It will be helpful to partly review general features of fluctuations
off the NJL chiral soliton in order to make the paper self--contained
and to allow for considerations which not only concern the S--wave channel
but are rather important for all channels. In section III we will
discuss the mutual relation of valence-- and sea--quark excitations
in contributing to the bound state. It is demonstrated that our
Bethe--Salpeter equation contains (unphysical) quark--antiquark
thresholds. These may be traced back to be deficiencies of the
non--confining NJL model. In section IV the numerical results for
$\Lambda$ (1405) will be presented. Finally, we will summarize and
conclude in section V. Furthermore two appendices are included
where we list relevant matrix elements and discuss the threshold
behavior of the NJL soliton model.

\medskip

\section{Kaon S--wave in the NJL soliton background}

\medskip
First, we wish to briefly review the treatment of meson
fluctuations off the NJL soliton. The starting point for our
explorations is given by the chirally symmetric NJL
Lagrangian for three flavors
\be
\La = \bar q (i\dslash - \hat m^0 ) q +
      2G_{\rm NJL} \sum _{i=0}^{8}
\left( (\bar q \frac {\lambda ^i}{2} q )^2
      +(\bar q \frac {\lambda ^i}{2} i\gamma _5 q )^2 \right) .
\label{NJL}
\ee
Here $\hat m^0$ represents the current quark mass matrix of the quark
field $q$. $\lambda ^i$ are the Gell--Mann matrices. We assume
isospin symmetry implying $m^0=m_u^0=m_d^0$. The NJL coupling
constant $G_{NJL}$ is determined from physical properties in
the baryon number zero sector of the model.

Application of functional integral bosonization techniques allows
to rewrite the action in terms of composite boson fields\cite{eb86}.
The NJL--action ${\cal A}_{\rm NJL}$ then is the sum of a purely
mesonic piece
\be
\A_m=\int d^4x\left(-\frac{1}{4G_{\rm NJL}}
\tr(M^{\dag}M-\hat m^0(M+M^{\dag})+(\hat m^0)^2)\right)
\label{ames}
\ee
and a fermion determinant
\be
\A_F=\Tr\log(i\Dslash)=
\Tr\log\left(i\dslash-(P_RM+P_LM^{\dag})\right),
\label{fdet}
\ee
\ie$\A_{NJL}=\A_F+\A_m$. The scalar and pseudoscalar fields are
parametrized by the complex matrix $M=S+iP$ while
$P_{R,L}=(1\pm \gamma _5)/2$ denote the right-- and left--handed
projectors.

The fermion determinant receives contributions from the valence
quark and the Dirac sea: $\A_F=\A_{\rm val}+\A_0$. In order to
regularize the ultra--violet divergent vacuum part of the action,
$\A_0$, we employ the Schwinger proper--time prescription\cite{sch51}
to the Wick rotated $(x_0=-i\tau)$ action in Euclidean space. Then the
vaccum part of the fermion determinant splits into real and imaginary
parts: $\A_0=\A_R+\A_I$. Proper--time regularization is realized by
introducing the $O(4)$ invariant cut--off $\Lambda$ via the
replacement
\be
\A_R=\frac{1}{2}\Tr\log\left(\Dslash_E^{\dag}\Dslash_E\right)
\longrightarrow -\frac{1}{2}\int_{1/\Lambda ^2}^\infty
\frac{ds}s\Tr\exp\left(-s\Dslash_E^{\dag}\Dslash_E\right) .
\label{arreg}
\ee
which is an identity for $\Lambda\rightarrow\infty$. Although the
imaginary part
\be
\A_I=\frac{1}{2}\Tr\log\left((\Dslash_E^{\dag})^{-1}\Dslash_E\right)
\label{af3}
\ee
is finite, in our calculations it will also be regularized in a
way consistent with eqn. (\ref{arreg}) (see below).

We parametrize the meson fields by\cite{we93a}
\be
M=\xi_0\xi_f\Sigma\xi_f\xi_0.
\label{defm}
\ee
where $\Sigma$ is a hermitian matrix. Spontaneous breaking of
chiral symmetry, which is a prominent feature of the NJL model,
is reflected by non--vanishing vacuum expectation
values $\langle \Sigma \rangle ={\rm diag}(m,m,m_s)$, the
entries of which are the constituent quark masses.
The unitary matrices $\xi_0$ and $\xi_f$ contain the information
on the static soliton and the meson fluctuations,
respectively. For the soliton we assume the hedgehog
{\it ansatz}
\be
\xi_0(\mbox {\boldmath $x $}) & = & {\rm exp}\left(\frac{i}{2}
{\mbox{\boldmath $\tau$}} \cdot{\bf \hat r}\ \Theta(r)\right)
\label{chsol}
\ee
with $\Theta$ being the chiral angle. In the baryon number zero sector
$\xi_0$ is substituted by the unit matrix. The pseudoscalar
fluctuations
\be
\xi_f(x)={\rm exp}\left(i\sum_{a=1}^8\eta_a(x)\lambda_a/2\right) .
\label{defeta}
\ee
are Fourier decomposed in Euclidean space
\be
\eta_a({\bf r},-i\tau) & = & \int_{-\infty}^{+\infty}
\frac{d\omega}{2\pi}\tilde \eta_a({\bf r},i\omega)
{\rm e}^{-i\omega\tau}.
\label{feta}
\ee
Here $\omega$ is the Euclidean frequency which after
regularization has to be continued back to Minkowski space in order
to make statements on physical observables.

For the ongoing discussion we will only allow
pseudoscalar meson fields to be space--time dependent. This
corresponds to replacing $\Sigma$ by its vacuum expectation
value in eqn. (\ref{defm}). Then the  Euclidean Dirac operator
$\Dslash_E$ may be written as
\be
i\beta\Dslash_E=-\partial_\tau-h=
-\partial_\tau-\mbox {\boldmath $\alpha \cdot p $}
-\T \beta\left(\xi_f\langle\Sigma\rangle\xi_f P_R
+\xi_f^\dagger\langle\Sigma\rangle\xi_f^\dagger P_L\right)\T^\dagger
\label{de1}
\ee
wherein $\tau=ix_0$ is the Euclidean time and the unitary matrix
\be
\T=\xi_0 P_L + \xi_0^\dagger P_R
\label{deft}
\ee
parametrizes the static soliton. We aim at expanding the
action up to quadratic order in the meson fluctuations
\be
\A_{\rm NJL} &=& \A^{(0)}+\A^{(1)}+\A^{(2)}+\ldots\ .
\label{expa}
\ee
The leading order term $\A^{(0)}$ corresponds to the action of the
static soliton\cite{re89}. This soliton minimizes $\A^{(0)}$\cite{re88}
enforcing the linear term $\A^{(1)}$ to vanish\cite{we93a}. The
important information on the fluctuations is contained in the quadratic part
$\A^{(2)}$. In order to obtain $\A^{(2)}$ also the expansion of the
Dirac Hamiltonian
\be
h=h_{(0)}+h_{(1)}+h_{(2)}+\ldots
\label{de2}
\ee
is required. In (\ref{de2}) the subscripts label the powers of
$\eta_a$. Obviously the zeroth order term only involves the chiral
angle and therefore is static
\be
\!\!h_{(0)}({\bf r})&=&\!\!\mbox {\boldmath $\alpha \cdot p $} +
{\cal T}\beta\langle\Sigma\rangle{\cal T}^\dagger =
{\mbox {\boldmath $\alpha \cdot p $}} + \beta
\pmatrix{m\ {\rm exp}\left(i\gamma_5{\mbox{\boldmath $\tau$}}
\cdot{\bf \hat r}\ \Theta(r)\right) & 0 \cr 0 & m_s\cr}.
\label{h0}
\ee
Since the chiral rotation ${\cal T}$ only involves static fields
the Fourier transformation for the fluctuation (\ref{defeta}) is
directly transferred to the perturbative terms in the Dirac Hamiltonian
\be
h_{(1)}({\bf r},-i\tau) & = & \int_{-\infty}^{+\infty}
\frac{d\omega}{2\pi}\tilde h_{(1)}({\bf r},i\omega)
{\rm e}^{-i\omega\tau} \nonumber \\*
h_{(2)}({\bf r},-i\tau) & = &
\int_{-\infty}^{+\infty}\frac{d\omega}{2\pi}
\int_{-\infty}^{+\infty}\frac{d\omega^\prime}{2\pi}
\tilde h_{(2)}({\bf r},i\omega,i\omega^\prime)
{\rm e}^{-i(\omega+\omega^\prime)\tau}.
\label{fham}
\ee
We are now equipped to perform the functional trace for general
fluctuations up to quadratic order in the soliton background.
The lengthy calculation has been carried out in ref. \cite{we93a}
and for completness we list the final result. As already noted the
fermion determinant receives contributions from the Dirac sea as well
the explicit occupation of the valence quark state. This, of course,
also applies to the quadratic part, \ie
$\A^{(2)}=\A^{(2)}_0+\A^{(2)}_{\rm val}$. Then we
have
\be
\A_0^{(2)} & = & \frac{N_C}{2}\int_{1/\Lambda^2}^\infty
\frac{ds}{\sqrt{4\pi s}}\sum_\mu 2\epsilon_\mu
{\rm e}^{-s\epsilon_\mu^2}\int^{+\infty}_{-\infty}
\frac{d\omega}{2\pi}\langle\mu|\tilde h_{(2)}({\bf r},\omega,-\omega)
|\mu\rangle
\nonumber \\*
&&\!\!\!+\frac{N_C}{4}\int_{1/\Lambda^2}^\infty ds \sqrt{\frac{s}{4\pi}}
\sum_{\mu\nu}\int^{+\infty}_{-\infty}\frac{d\omega}{2\pi}
\langle\mu|\tilde h_{(1)}({\bf r},\omega)|\nu\rangle
\langle\nu|\tilde h_{(1)}({\bf r},-\omega)|\mu\rangle
\label{afquad}
\\*
&&\times \left\{\frac{{\rm e}^{-s\epsilon_\mu^2}
+{\rm e}^{-s\epsilon_\nu^2}}{s}
+[\omega^2-(\epsilon_\mu+\epsilon_\nu)^2]
R_0(s;\omega,\epsilon_\mu,\epsilon_\nu)
-4\omega\epsilon_\nu R_1(s;\omega,\epsilon_\mu,\epsilon_\nu)
\right\}
\nonumber
\ee
with $|\mu\rangle$ being eigenstates of the zeroth order
Hamiltonian $h_{(0)}({\bf r})$ possessing eigenvalues $\epsilon_\mu$.
The regulator functions contain Feynman parameter integrals
\be
R_i(s;\omega,\epsilon_\mu,\epsilon_\nu)
=\int_0^1 x^i dx\ {\rm exp}\left(-s[(1-x)\epsilon_\mu^2
+x\epsilon_\nu^2-x(1-x)\omega^2]\right)
\label{regfct}
\ee
reflecting the quark loop. $s$ denotes the proper--time parameter
introduced in the regularization (\ref{arreg}).  The contributions
to (\ref{afquad}) which are of odd powers in $\omega$ stem from
the imaginary part of the action (\ref{af3}). For these terms
proper--time regularization has been imposed by introducing
a parameter integral for the propagator in the quark loop
\cite{we93a,al92b}. Although the expression for $\A_{\rm val}$ may
be found in refs. \cite{we93a,we93b} we present it here, because it
will be of relevance for the later discussion on the applicability
of the model,
\be
\A_{\rm val}^{(2)} & = & -\eta^{\rm val}N_C
\int^{+\infty}_{-\infty}\frac{d\omega}{2\pi}
\Big(\langle{\rm val}|\tilde h_{(2)}({\bf r},\omega,-\omega)
|{\rm val}\rangle \nonumber \\*
&&\qquad\qquad\qquad
+\sum_{\mu\ne{\rm val}}\frac{
\langle{\rm val}|\tilde h_{(1)}({\bf r},\omega)|\mu\rangle
\langle\mu|\tilde h_{(1)}({\bf r},-\omega)|{\rm val}\rangle}
{\epsilon_{\rm val}-\omega-\epsilon_\mu}\Big) .
\label{valquad}
\ee
Here $\eta^{\rm val}=0,1$ denotes the occupation number of the
valence quark and anti-quark states. $|{\rm val}\rangle$ refers to
the valence quark eigenstate of $h_{(0)}$. Finally, the contribution
of the purely mesonic part (\ref{ames}) has to be included. It turns
out to be of the same analytical form as for the P--wave.

The central issue in ref. \cite{we93b} has been to explore the
kaon bound state in the P--wave channel. It is thus the main goal here
to work out the special form of the perturbative parts $\tilde h_{(1)}$
and $\tilde h_{(2)}$ for kaon fluctuations and the associated matrix
elements $\langle\mu|\tilde h_{(1)}|\nu\rangle$ in the S--wave channel.
This confines the fluctuations to the subspace
\be
\sum_{\alpha=4}^7\tilde\eta_\alpha({\bf r},\omega )\lambda_\alpha=
\pmatrix{0 & \tilde K({\bf r},\omega)\cr
\tilde K^\dagger({\bf r},-\omega)&0\cr}
\label{stfluc}
\ee
wherein $\tilde K({\bf r},\omega)$ is a two--component isospinor. For
the S--wave these two components are pure radial functions\cite{da89}
\be
\tilde K({\bf r},\omega)=\pmatrix{a(r,\omega) \cr b(r,\omega)\cr}.
\label{bound}
\ee
Substituting the expressions (\ref{stfluc},\ref{bound}) the
perturbative terms in the Dirac Hamiltonian are obtained to be
\be
\!\!\!\tilde h_{(1)}({\bf r},\omega)&=&-\frac{1}{2}
(m+m_s)\pmatrix{0 & u_0({\bf r})\tilde K(r,\omega)\cr
\tilde K^\dagger(r,-\omega) u_0({\bf r})&0\cr}
\label{h1str} \\
\nonumber \\*
\!\!\tilde h_{(2)}({\bf r},\omega,-\omega)
&=&\frac{1}{4}(m+m_s)
\pmatrix{u_0({\bf r})\beta\tilde K(r,\omega)
\tilde K^\dagger(r,\omega)u_0({\bf r}) & 0 \cr
0& \hspace{-1cm}
-\beta\tilde K^\dagger(r,-\omega)\tilde K(r,-\omega)\cr}.
\label{h2str}
\ee
For convenience we have introduced the unitary, self-adjoint
transformation matrix $u_0$
\be
u_0({\bf r})=\beta\left({\bf \hat r}\cdot{\mbox{\boldmath $\tau$}}
{\rm sin}\frac{\Theta}{2} -i\gamma_5{\rm cos}\frac{\Theta}{2}\right),
\label{u0}
\ee
which obviously is odd under parity transformations. Therefore
$\tilde h_{(1)}$ couples states of opposite parity in the matrix
elements $\langle\mu|\tilde h_{(1)}|\nu\rangle$. The main
difference with respect to the P--wave kaon is an overall factor
${\bf \hat r}\cdot {\mbox{\boldmath $\tau$}}$ for $u_0({\bf r})$
\cite{we93b}. In that case $\tilde h_{(1)}$ is therefore of even
parity and only connects states of identical parity. Since the
regulator functions for S-- and P--waves are identical the main
effort of algebraic manipulations goes into the evaluation of the
matrix elements which involve parity changes. It turns out that the
matrix elements of $\tilde h_{(2)}$ are of the same analytical
structure as for the P--wave \cite{we93b}. The S--wave matrix elements
for $\langle\mu|\tilde h_{(1)}|\nu\rangle$ are given in appendix
\ref{matrix}. Then the action functional for the fluctuations is
given as a bilocal integral for $a(r,\omega)$ and $b(r,\omega)$ defined
in eqn. (\ref{bound}). Due to isospin invariance the radial dependence of
$a(r,\omega)$ and $b(r,\omega)$ is identical and we may introduce
Fourier amplitudes $a_i(\omega)$
\be
a(r,\omega)=\eta_\omega(r)a_1(\omega)
\qquad {\rm and}\qquad
b(r,\omega)=\eta_\omega(r)a_2(\omega).
\label{annihil}
\ee
The radial function $\eta_\omega(r)$ is found as the solution to
the variation of the quadratic piece  $\delta\A^{(2)}/\delta\eta=0$
which turns out to be a homogeneous linear integral equation and
may be cast into the general form
\be
r^2\left\{\int dr^\prime r^{\prime2} \Phi_S^{(2)}(\omega;r,r^\prime)
\eta_\omega(r^\prime)+\Phi_S^{(1)}(\omega;r)\eta_\omega(r)
\right\}=0.
\label{eqmfluc}
\ee
The subscript $S$ has been attached to emphasize that these
(bi-)local kernels refer to S--wave kaons in the soliton background.
These kernels are obtained from the analogous expression for the
P--wave in ref.\cite{we93b} by substituting the S--wave $u_0({\bf r})$
defined in eqn. (\ref{u0}). In fact, eqn. (\ref{eqmfluc}) represents
the Bethe--Salpeter equation for S--wave kaons in the soliton
background.

In order to project onto states which carry the quantum numbers of real
baryons we apply the semiclassical zero--mode quantization\cite{ad83}.
Time dependent collective coordinates describing the isospin
orientation are therefore introduced via
\be
M=R(t)\xi_0\xi_f\langle\Sigma\rangle\xi_f\xi_0R^\dagger(t)
\qquad {\rm with}\qquad R(t)\in SU(2).
\label{crank}
\ee
The time dependence is most conveniently made transparent by
defining angular velocities
\be
\frac{i}{2}{\mbox {\boldmath $\tau$}}\cdot
{\mbox {\boldmath $\Omega$}}=R^\dagger(t){\dot R}(t).
\label{velocity}
\ee
In addition to the kaon fluctuations the action is also expanded
in ${\mbox {\boldmath $\Omega$}}$. Assuming the existence of a
bound state with eigenenergy $\omega_0$ in the kaon S--wave
channel this leads to the collective Lagrangian involving the
collective rotations
\be
L_{\mbox{\boldmath $\Omega$}}=
\frac{1}{2}\alpha^2{\mbox{\boldmath $\Omega$}}^2
-\frac{1}{2}  c(\omega_0)
{\mbox{\boldmath $\Omega$}}\cdot
\left(\sum_{i,j=1}^2 a_i^\dagger(\omega_0)
{\mbox{\boldmath $\tau$}}_{ij}
a_j(\omega_0)\right)+\cdot\cdot\cdot.
\label{lcoll}
\ee
The dots represent terms of order ${\mbox{\boldmath $\Omega$}}^3$
as well as contributions form the meson continuum\footnote{Of course,
also the contribution from the P--wave bound state should appear
in (\ref{lcoll}). However, we will only be concerned with the
S--wave channel here.}. The moment of inertia $\alpha^2$ is
discussed at length in the literature \cite{re89,go91}.
The analytical expression for $c(\omega_0)$ is the same as for
the P--wave, however, the transformation matrix (\ref{u0})
as well as the solution $\eta_{\omega_0}(r)$ to the S--wave
Bethe--Salpeter equation have to be substituted in eqn. (4.6) of
ref. \cite{we93b}. The arbitrary
normalization of $\eta_{\omega_0}(r)$ which transfers to $c(\omega_0)$
may be eliminated by noting that the bound state contribution to
the spin\footnote{In ref. \cite{we93b} is has been shown that the
spin carried by the kaon fluctuations is identical to the
expectation value of the grand spin.} \ involves the same
dependence on the Fourier amplitudes $a_i(\omega_0)$
as the last term in (\ref{lcoll})
\be
{\bf J}_K = - \frac{1}{2}  d(\omega_0)
\left(\sum_{i,j=1}^2 a_i^\dagger(\omega_0)
{\mbox{\boldmath $\tau$}}_{ij}
a_j(\omega_0)\right)+({\rm meson\ continuum}).
\label{jk}
\ee
The total spin is the sum of ${\bf J}_K$ and the spin carried by the
soliton ${\bf J}_\Theta= \partial L_{\mbox {\boldmath $\Omega$}}/
\partial{\mbox {\boldmath $\Omega$}}$. The hedgehog structure of the
soliton yields the identity $|{\bf J}_\Theta|=|{\bf I}|$ where
${\bf I}$ is the total isospin\cite{ca85}.  We have now collected all
ingredients for the baryon mass formula which is derived from
(\ref{lcoll}) analogously to the P--wave formalism\cite{ca85,da89}
\be
M_B=E_{\rm cl}+S\omega_0+\frac{1}{2\alpha^2}
\left(\chi J(J+1)+(1-\chi)I(I+1)\right).
\label{mass}
\ee
Here $S$ denotes the strangeness charge of the baryon under
consideration. Obviously, $\chi=-c(\omega_0)/d(\omega_0)$ is
independent of the normalization of $\eta_{\omega_0}(r)$.

\medskip

\section{Interplay between valence-- and sea--quark
mesonic excitations}

\medskip

In this section we discuss the principle behavior of the solution
to the bound state equation (\ref{eqmfluc}). In purely mesonic
soliton models like the Skyrme model, the bound state equation
is essentially a Klein--Gordon equation for pseudoscalar mesons
in an external localized field which is generated by the static
soliton\cite{ca85,da89}. Then there appear kaon bound states in this
background field possessing eigenenergies $\omega_0^2<m_K^2$ below
the kaon threshold $E_{\rm th}=m_K$  for the continuum of unbound
kaons.

In the presently considered bosonized NJL model the bound state
equation is given by the homogeneous integral equation (\ref{eqmfluc})
for the radial functions $\eta_\omega(r)$. It is important to note
that the kernels of this integral equation are expressed as mode sums
involving the eigenstates of the one--particle Hamiltonian
$h_{(0)}$ (\ref{h0}). Accordingly, the bound state is determined
by the structure of the single quark spectrum. Apart form
relativistic covariance and regularization the bound state equation
in the NJL model is very similar to the RPA--equation for finite
nuclei\footnote{For a field theoretical formulation of the RPA see
for example refs.\cite{re78}.}. It is therefore instructive to discuss
the spectrum of the one--particle Hamiltonian $h_{(0)}$ (\ref{h0}).
Obviously the strange quark orbits are not effected by the presence
of the chiral soliton. Moreover, the non--strange quark levels, in
particular those of the Dirac sea, are only slightly distorted by
the self--consistent soliton field except for the lowest lying
$G^\pi=0^+$ valence quark state $|{\rm val}\rangle$ with eigenvalue
$\epsilon_{\rm val}$ which is strongly bound as $m$ increases.
Here $G$ denotes the eigenvalue of the grand spin operator
${\bf G}={\bf J}+ {\mbox{\boldmath $\tau$}}/2$ which commutes with
$h_{(0)}$. The qualitative form of the quark spectrum is displayed
in Fig. 1.

\begin{figure}
\begin{center}
\setlength{\unitlength}{1mm}
\begin{picture}(150,91)
\put(25,50){\line(1,0){90}}
\thicklines
\put(134,68){$E_{\rm th}^{\rm val}$}
\put(130,68){\line(1,0){1.2}}
\put(130,60){\line(0,1){16}}
\put(128,60){\line(1,0){2.1}}
\put(128,76){\line(1,0){2.1}}
\put(25,60){\line(1,0){30}}
\put(25,60.2){\line(1,0){30}}
\put(60,60){$\epsilon_{\rm val}$}
\put(25,70){\line(1,0){30}}
\put(25,70.2){\line(1,0){30}}
\put(60,70){$m$}
\put(25,30){\line(1,0){30}}
\put(25,29.8){\line(1,0){30}}
\put(60,29){$-m$}
\multiput(40,60.5)(6,2){7}{\line(3,1){4}}
\put(82,74.5){\line(3,1){5.2}}
\multiput(50,30)(6,6){8}{\line(1,1){4}}
\put(25,10){\line(1,0){30}}
\put(25,12){\line(1,0){30}}
\put(25,14){\line(1,0){30}}
\put(25,16){\line(1,0){30}}
\put(25,18){\line(1,0){30}}
\put(25,20){\line(1,0){30}}
\put(25,22){\line(1,0){30}}
\put(25,24){\line(1,0){30}}
\put(25,26){\line(1,0){30}}
\put(25,28){\line(1,0){30}}
\put(25,72){\line(1,0){30}}
\put(25,74){\line(1,0){30}}
\put(25,76){\line(1,0){30}}
\put(25,78){\line(1,0){30}}
\put(25,80){\line(1,0){30}}
\put(25,82){\line(1,0){30}}
\put(25,84){\line(1,0){30}}
\put(25,86){\line(1,0){30}}
\put(25,88){\line(1,0){30}}
\put(25,90){\line(1,0){30}}
\put(120,76){$m_s$}
\put(120,49){0}
\put(120,23){$-m_s$}
\put(85,10){\line(1,0){30}}
\put(85,12){\line(1,0){30}}
\put(85,14){\line(1,0){30}}
\put(85,16){\line(1,0){30}}
\put(85,18){\line(1,0){30}}
\put(85,20){\line(1,0){30}}
\put(85,22){\line(1,0){30}}
\put(85,24){\line(1,0){30}}
\put(85,23.8){\line(1,0){30}}
\put(85,76.2){\line(1,0){30}}
\put(85,76){\line(1,0){30}}
\put(85,78){\line(1,0){30}}
\put(85,80){\line(1,0){30}}
\put(85,82){\line(1,0){30}}
\put(85,84){\line(1,0){30}}
\put(85,86){\line(1,0){30}}
\put(85,88){\line(1,0){30}}
\put(85,90){\line(1,0){30}}
\end{picture}
\end{center}
\noindent
\fcaption{A schematic plot of the spectrum of the one--particle
Hamiltonian (\ref{h0}). We distinguish between non--strange (left)
and strange (right) states. The dashed lines indicate possible kaonic
excitations.}
\end{figure}

The kaon fluctuations $\tilde K$ are built up from $s\bar q (q=u,d)$
excitations. In the vacuum the free kaon is generated by a
coherent sum of $s\bar q$ excitations of the Dirac sea (see
also Fig. 1). Due to the lack of confinement in the NJL model
there is an $s\bar q$ threshold starting at $E_{\rm th}\approx
m+m_s$. In a confining theory quarks do not exist as asymptotic
(free) states and this threshold would thus be absent.
In the presence of the self--consistent soliton field (with the $0^+$
valence orbit occupied) the kaonic fluctuation mode receives in addition
an $s \bar q$ contribution from the excitation of a non-strange-quark
in the valence quark orbit $G^\pi = 0^+$ to a strange quark, see Fig. 1.
Due to the occupation of $|{\rm val}\rangle$ in the presence of the
soliton the $s \bar q$ threshold is lowered to $E^{val}_{th}=
m_s-m+\triangle E$ where $\triangle E=m-\epsilon_{\rm val}$ is the
binding energy of the occupied $G^\pi = 0^+$ orbit. From a formal
point of view this threshold is manifested by a singularity in the
expression for the valence quarks' contribution to the action
(\ref{valquad}). Note that the denominator in the second expression
on the $RHS$ of this equation vanishes for $\omega=\epsilon_{\rm val}
-\epsilon_\mu$, where $\epsilon_\mu$ is an eigenvalue of a state in
the strange Dirac sea. The threshold is thus determined by the
lowest eigenvalue $\epsilon_\mu\approx m_S$.
For $\omega<-E^{val}_{th}$ the Dirac equation
for the strange quarks allows oscillating (\ie free) solutions.
As seen from Fig. 2 the $s \bar q$ threshold due to the occupation
of the valence quark level is considerably lower than the one in the
vacuum $E_{th} \simeq m+m_s$ and represents the upper bound for possible
kaonic bound states in the soliton background. In Fig. 2 the strange
constituent quark mass $m_s$ and the energy of the $G^\pi = 0^+$
(non-strange) valence quark state are shown as functions of the
non-strange constituent quark mass $m$. It is seen that this threshold
increases almost linearly with $m$.
\begin{figure}
\centerline{
\psfig{figure=thres.tex.ps,height=9cm,width=9cm}
\hfil
\hspace{-1cm}
\psfig{figure=gap.tex.ps,height=9cm,width=9cm}
}
\fcaption{Left: The threshold behavior
in the presence of the soliton. The solid line refers to
the actual threshold while the dashed line denotes the
threshold in the absence of the valence quark.
Right: The valence quark level $\epsilon_{\rm val}$ (solid line) and
the continuum for the strange quarks (shaded area). The lower
bound of the ``positive" part of the continuum is $m_s$ and the
``negative" part is bounded from above by $-m_s$.}
\end{figure}
\noindent
For sufficiently large $m$ the occupied (non-strange) valence quark
orbit dives into the Dirac sea $\epsilon_{\rm val}  < 0$ and there is no
explicit valence quark. Consequently the valence quark $s \bar q$
threshold $E^{val}_{th}$ is abandoned. The valence quark is included
in the vacuum (Dirac sea) and at $\epsilon_{\rm val} = 0$ the valence
threshold matches continuously into the threshold for $s \bar q$
excitations from the distorted Dirac sea
$E^{sea}_{th} = \vert \epsilon_{\rm val} \vert +m$ ({\it cf.} Fig. 2).
This threshold is obtained by considering the Feynman parameter integral
(\ref{regfct}). The threshold shows up when $\omega$ is chosen such
that the argument of the exponential acquires exactly one real root in
the interval $0\le x \le 1$. The reader may consult
appendix \ref{threshold} for details on this calculation.

In order to numerically solve the Bethe--Salpeter equation
(\ref{eqmfluc}) it is discretized on a one--dimensional lattice
$r_i=i\triangle r$ and extended to an eigenvalue problem
\be
r_i^2\sum_j\left\{\triangle r\ r_j^2\ \Phi^{(2)}_{ij}(\omega)
+\delta_{ij}\Phi^{(1)}_{j}(\omega)\right\}\eta_\omega(j)
=\lambda(\omega) \eta_\omega(i).
\label{eigprob}
\ee
The solution to the Bethe--Salpeter equation obviously is obtained
for $\lambda(\omega_0)=0$. The above discussed threshold behavior
also shows up in the dispersion of the eigenvalues $\lambda(\omega)$.
This is demonstrated in Fig. 3 where the eigenvalues with the
lowest absolute values are plotted as functions of the frequency
$\omega$.
\begin{figure}
\centerline{
\psfig{figure=eig.430.tex.ps,height=9cm,width=9cm}
\hfill
\hspace{-1cm}
\psfig{figure=eig.550.tex.ps,height=9cm,width=9cm}
}
\fcaption{The lowest eigenvalues of the Bethe--Salpeter kernel
(\ref{eigprob}) as functions of the Fourier frequency $\omega$
in the vicinity of the bound state energy. Two specific values
of the up constituent mass $m$ are displayed.}
\end{figure}
In the case $m=550$MeV we obtain $E^{val}_{th}=600$MeV ({\it cf.}
Fig. 2) which is significantly larger than the kaon mass
$m_K=495$MeV. At $\omega_0=-459$MeV one of the eigenvalues
smoothly goes through zero and a solution to the Bethe--Salpeter
equation can be identified. On the contrary, for $m=430$MeV
$E^{val}_{th}$ appears at 442MeV$<m_K$. The vanishing denominator in
eqn. (\ref{valquad}) then causes some eigenvalues $\lambda(\omega)$
to diverge for $\omega<-E^{val}_{th}$. Since we are dealing with
discretized eigenenergies of $h_{(0)}$ (\ref{h0}) these singularities
are also distinct. This discretization is achieved by inserting the
system in a large (compared to the extension of the soliton) spherical
box of radius $D$ and imposing boundary conditions\footnote{For the
discussion of various boundary conditions see \eg ref.\cite{we92}. Our
boundary condition are such that the upper components of the spinors
vanish at $r=D$.} on the eigenfunctions of $h_{(0)}$. A further
consequence of this discretization is the fact that the first
singularity for $\lambda(\omega)$ does not exactly appear at
$-E^{val}_{th}$ but rather at $-\sqrt{(E^{val}_{th})^2+(q_1^1/D)^2}$.
Here $q_1^1\approx4.49$ denotes the smallest root of the Bessel
function $j_1$ associated with orbital angular momentum $l=1$.
In the continuum limit, of course, these singularities are dense.
Each of these singularities correspond to oscillating strange
quark solutions of the Dirac equation which also contains the kaon
fluctuation, \ie scattering solutions. Thus these singularities are
due to the lack of confinement in the NJL model. These divergences
are also accompanied by a ``no--crossing'' phenomena for the other
eigenvalues.

\medskip

\section{Numerical results}

\medskip

For the numerical calculations the parameters are fixed in the baryon
number zero (meson) sector of the model. Since the Schwinger--Dyson
equations allow to express the current quark masses $m^0_i$ in terms
of the constituent quark masses $m_i$ and the cut--off $\Lambda$, the
model contains four undetermined parameters. Assuming the physical
values for the pion decay constant $f_\pi=93$MeV as well as for the
pion and kaon masses $m_\pi=135$ and $m_K=495$MeV reduces the number
of unknown parameters to only one which we choose to be the constituent
mass of the non--strange quark: $m$.  The corresponding results for the
parameters $m^0_i, m_s, G_{\rm NJL}$ and $\Lambda$ as well as the
corresponding prediction for the kaon decay constant obtained by the use
of the proper--time regularization may \eg be found in ref.\cite{we92}.
For a given value of $m$ we construct the soliton which minimizes the
static part of the action, $\A^{(0)}$ in eqn. (\ref{expa}),
self--consistently. For details on the iterative procedure used for
this calculation the reader may consult the literature where this
technique has been discussed extensively\cite{re88}. The resulting
profile function of the chiral angle, $\Theta(r)$, as well as the
eigenvalues, $\epsilon_\mu$, and eigenstates, $|\mu\rangle$, of the
static Hamiltonian $h_{(0)}$ (\ref{h0}) are then employed to evaluate
the integral kernels $\Phi_S^{(1)}$ and $\Phi_S^{(2)}$ of the
Bethe--Salpeter equation (\ref{eqmfluc}). As explained in the previous
section the Bethe--Salpeter equation is extended to an eigenvalue
problem. The frequency $\omega$ is then adjusted to yield a vanishing
eigenvalue $\lambda$ in eqn. (\ref{eigprob}). $\omega$ is interpreted
as the bound state energy and the corresponding eigenvector as the
(unnormalized) bound state wave--function.

In order to test our numerical routines we have considered the
low--lying bound state solutions of the Bethe--Salpeter equation
for a vanishing chiral angle, \ie the baryon number zero case.
In the special case $D=6$fm we have found eigenenergies
$\omega_0=\pm511,\pm546\ {\rm and}\ \pm596$MeV. Assuming the
$D$--dependence $\omega_0^2=\tilde m_K^2+(n\pi/D)^2$ these numbers
can be fitted with $\tilde m_K\approx504$MeV for $n=1,2$ and $3$.
We have observed the same behavior for $D=5$fm. Of course, in the
case of baryon number zero we should have $\tilde m_K=m_K$. This
demonstrates that our numerical results are trustworthy to the order
of 2\%. We should also mention that the ``bound state" wave--functions
for these three cases compare reasonably well with the Bessel
functions $j_0(n\pi r/D)$ which describe the radial behavior of an
S--wave. For the bound state in the soliton background we expect an
even better accuracy than the above mentioned 2\% since due to their
localization they are less sensitive to $1/D$ effects. This statement
is supported by the fact that in the SU(3) symmetric case ($m_\pi=m_K$)
our numerical procedure reproduces the corresponding P--wave
zero--mode with good accuracy at $\omega=-0.2$MeV \cite{we93b}.

The discussion in the previous section has shown that in the presence
of the soliton a quark--antiquark channels opens below $m_K$ when
non--strange constituent masses less than $m=450$MeV are considered.
{}From Skyrme model results\cite{da89} we expect the S--wave bound state
to be only losely bound \ie the absolute value of the eigenenergy
$\omega_0$ being only a few MeV less than $m_K$. Therefore we cannot
make definite predictions on the S--wave for $m<450$MeV in the
framework of the NJL model. This is somewhat unfortunate since the NJL
model is known to reasonably well reproduce light baryon properties
for the parameter region $400{\rm MeV}\le m\le 450{\rm MeV}$\cite{wa91}.
{\it E.g.} the empirical value for the nucleon $\Delta$ mass
difference (293MeV) is obtained for $m\approx430$MeV\cite{go91}.
For $m>430$MeV the NJL model overestimates this mass difference.

As will be demonstrated, we in fact do find S--wave kaon bound states in
the soliton background for $m\ge450$MeV. The corresponding bound state
wave functions $\eta(r)$ are displayed in Fig. 4. These wave
functions are normalized to carry strangeness $S=-1$. The strangeness
charge carried by the kaon mode may be evaluated similarly to the
P--wave case ({\it cf.} eqn. (3.18) in ref. \cite{we93b}) by substituting
the corresponding expressions for $u_0({\bf r})$ and $\eta_{\omega_0}$.
For $\Theta=0$ we find that states with $\omega_0>0$ have $S>0$ while
states with $\omega_0<0$ carry negative strangeness\footnote{A strange
quark of positive energy is defined to have $S=-1$.}. Also for the
bound state solutions with $-m_K<\omega_0<0$ the calculated strangeness
charge is negative. Thus these states may be normalized to $S=-1$.
In the presence of the soliton the degeneracy between states with
$\pm\omega_0$ is removed by the imaginary part of the action $\A_I$.
In Skyrme type models $\A_I$ is mocked up by the Wess--Zumino term
\cite{wi83} which is the leading term in the gradient expansion of
$\A_I$. Obviously the soliton enforces the bound state
wave--function to vanish at the origin in contrast to free S--waves
which are finite at $r=0$. The shape of the wave--functions is similar
to the results found in ref. \cite{sc88} for the Skyrme model. Up to
normalization their parametrization of the kaon fluctuation is identical
to ours (\ref{defm}). Finally we remark that we do not find S--wave
bound states with $\omega_0>0$ which would correspond to exotic
baryons with $S=+1$.

\begin{figure}
\centerline{\hskip -1.5cm
\psfig{figure=profile.tex.ps,height=9.0cm,width=16.0cm}}
\fcaption{The kaon bound state S-wave for several values of the
constituent quark mass $m$. Also shown is a solution to the
Bethe--Salpeter equation in the absence of the soliton $(\Theta=0)$.
In that case we have used D=5fm and obtained $\omega_0=\pm$518.3MeV.
The profile functions are normalized to carry strangeness S=-1.}
\end{figure}

\begin{table}
\tcaption{Parameters for the baryon mass formula
(\ref{mass}) and the prediction for the lightest S--wave
$\Lambda$ state relative to the nucleon $M_{\Lambda_S}-M_N$ as
functions of the constituent mass $m$. Also the resulting
values for the $\Delta$ nucleon mass splitting are presented.}
{}~
\newline
\centerline{
\begin{tabular}{|c|c|c|c|c|c|}
   \hline
$m$ (MeV) & $\alpha^2$ (1/GeV) & $\omega_0$ (MeV) & $\chi$ &
$\Lambda_S-N$ (MeV) &$\Delta-N$ (MeV) \\
\hline
450 & 4.78 & -461 & 0.46 & 418 & 314 \\
\hline
500 & 4.19 & -461 & 0.54 & 419 & 358 \\
\hline
550 & 3.74 & -459 & 0.57 & 416 & 401 \\
\hline
600 & 3.42 & -456 & 0.60 & 411 & 439 \\
\hline
\end{tabular}}
\end{table}

In table 1 we present the ingredients of the baryon mass
formula (\ref{mass}) as functions of the constituent quark
mass $m$. The bound state energy $\omega_0$ exhibits only
a very moderate dependence on $m$ while $\chi=-c(\omega_0)/
d(\omega_0)$ increases significantly with $m$. For $m=450$MeV
the quark--antiquark threshold $-E^{val}_{th}=-468$MeV
is slightly larger (in magnitude) than $\omega_0=-461$MeV for
which we find a solution to the Bethe--Salpeter equation. We
may therefore conclude that this solution indeed corresponds to
a kaon bound state. We have also verified that these results are not
sensitive to the chosen value of the box--radius $D$. {\it E. g.}
for the constituent mass $m=450$MeV our result for $\omega_0$
increases from -461.5MeV to -460.8MeV when changing $D$ from 5fm
to 6fm. Simultaneously $\chi$ increases also only sightly from 0.45
to 0.46. This moderate dependence on $D$ is due to the localization
of the bound state, making ``observables" less sensitive to the
large distance behavior of the kernels like $\Phi_S^{(2)}$ than
in the case of free kaons. It is interesting to note that for positive
valence quark energies the valence quark and sea quark parts of the
action contribute equally to both $c(\omega_0)$ as well as $d(\omega_0)$.
This is in contrast to the results for the P--wave where these
quantities are dominated by their valence quark
parts. Although the coupling $c(\omega_0)$ comes out to be larger
in magnitude for the S--wave than for the P--wave the parameter
$\chi$ which determines the coupling of the collective rotations
to the bound state turns out to be smaller. This contrast to the
Skyrme model may be understood by noting that the amplitude of the
spin carried by the bound kaon $d(\omega_0)$ is unity in the Skyrme
model since no states with grand spin larger than $1/2$ are involved
there\cite{ca85}. In the NJL model, however, the evaluation of the
fermion determinant enforces the consideration of states with arbitrary
grand spin. As it turns out for the S--wave $d(\omega_0)$ is slightly
larger than unity while for the P--wave
$d(\omega_0)\approx0.9$\cite{we93b}.

Finally, table 1 also contains our main numerical result: the
prediction for the mass difference between the lowest strange S--wave
baryon and the nucleon. The experimental value for this quantity is
466MeV. Obviously the bound state energy $\omega_0$ reproduces this
number perfectly, however, the prediction for the mass difference
is about 70MeV too low due to projection onto states with
good spin and isospin. From the resulting $\Delta$--nucleon
mass difference we conclude that the model makes reliable
predictions only for constituent quark mass up to about 500MeV
yielding an odd parity $\Lambda$ state which is 419MeV heavier than
the nucleon. Obviously this state is stable with respect to the decay
into a nucleon and a kaon although it might decay into a $\Sigma$ and
a pion. In our calculations the latter channel, however, is not open
since we only admit strange fluctuations, {\it cf}. eqn. (\ref{stfluc}).
The mass formula (\ref{mass}) also allows to make predictions on other
strange baryons of odd parity. For example, a $\Sigma$--state with a
mass of 534MeV+$M_N$ is obtained when using $m=450$MeV. However, such
a state is not stable with respect to a decay into a nucleon and a kaon.

\medskip

\section{Conclusions}

\medskip

In this paper we have investigated the structure of the kaonic S--wave
channel in the background field of the NJL chiral soliton. This has
been a straightforward but nevertheless fruitful application of a
formalism which allows to explore meson fluctuations off the NJL
soliton.  We have observed that the NJL model indeed allows for bound
states in this channel, \ie solutions of the Bethe--Salpeter
equation (\ref{eqmfluc}) with eigenenergies smaller than the
kaon mass. We have seen that the presence of the soliton and the
lack of confinement cause an additional and unphysical threshold
to pop up, which for some choices of the constituent quark mass $m$
is lower than the kaon mass. Unfortunately this is also the
case for commonly accepted values of $m$ which provide reasonable
predictions in the non--strange sector of the model\cite{wa91}.
As has been demonstrated in ref. \cite{we93b} treating the NJL model
within the bound state approach gives a suitable description of the
low--lying P--wave baryons for $m\simeq430$MeV. Also treating the
NJL model within the collective approach together with the exact
diagonalization of the symmetry breaking Hamiltonian\cite{ya88}
yields reasonable results for the mass spectrum of the even parity
hyperons\cite{we92,bl93} when considering $400\le m\le 450$MeV. This
approach, however, does not provide a description of the odd parity
$\Lambda(1405)$. The presently obtained results suggest that there
is only a narrow margin in parameter space 450MeV$\le m\le$500MeV where
the NJL model should be taken seriously, at least for dynamical
calculations which involve time dependent meson fields. For these
parameters the NJL model predicts an odd--parity $\Lambda$ state to
be stable against the decay into a nucleon and a kaon. The mass of
this state is found to be 419MeV higher than that of the nucleon. This
compares well with the $\Lambda(1405)$ which has a mass difference to
the nucleon of 466MeV.

The question of how to cure the problem of the undesired
quark--antiquark threshold arises naturally. Since presently
no method is known how to extend the NJL model to incorporate
confinement without loosing much of its feasibility one might
nevertheless wonder which kind of extensions could provide a
remedy of this problem in the sense that this threshold gets
shifted to energies high enough not to be relevant for the quantities
under consideration. One way in this direction \eg could be the inclusion
of vector mesons. It is confirmed\cite{al92} that then the valence
quark energy is much lower than in the presently considered model
and might even dive into the Dirac sea (\ie becomes part of the vacuum).
In that case the threshold for quark--antiquark excitations would at
least be as large as the strange constituent quark mass and thus
larger than the kaon mass. Treating, however, fluctuating strange
vector mesons consistently appears to already be rather involved in
the Skyrme model, especially for determining the coupling of these
fluctuations to the collective iso--rotations\cite{sc88}. Therefore we
anticipate that such calculations in the framework of the NJL model
soliton are exceedingly difficult.

\medskip

\setcounter{equation}{0}
\renewcommand{\theequation}{A.\arabic{equation}}
\appendix

\section*{Appendix A: Matrix elements of ${h_{(1)}}$} \label{matrix}

\medskip

In this appendix we present the matrix elements of $\langle \mu
|h_{(1)}({\bf r},\omega) |\rho\rangle$ for an S--wave kaon
({\it cf.} eqn. (\ref{h1str})). Both $|\mu\rangle$ and $|\rho
\rangle$ are eigenstates of the static Hamiltonian
$h_{(0)}({\bf r})$ (\ref{h0}) and differ by one unit of
strangeness.

First we construct eigenstates of the grand spin operator
${\bf G}={\bf J}+{\mbox{\boldmath $\tau$}}/2$ which commutes
with $h_{(0)}({\bf r})$. These states are characterized by their
grand spin quantum numbers $G$ and the corresponding projection $M$
as well as by spin and orbital angular momentum quantum numbers
$j$ and $l$ respectively: $|ljGM\rangle$. Then the eigenfunctions
$\Psi_\mu^{(G \pm)}$ of $h_{(0)}({\bf r})=
\langle {\bf r}|\mu G M \pm \rangle$ are of the
form
\be
\Psi_\mu^{(G,+)}=
\pmatrix{ig_\mu^{(G,+;1)}(r)|GG+\frac{1}{2}GM\rangle \cr
f_\mu^{(G,+;1)}(r) |G+1G+\frac{1}{2}GM\rangle \cr} +
\pmatrix{ig_\mu^{(G,+;2)}(r)|GG-\frac{1}{2}GM\rangle \cr
-f_\mu^{(G,+;2)}(r) |G-1G-\frac{1}{2}GM\rangle \cr}
\label{psipos}
\ee
for states with parity $\Pi=(-)^G$. For the parity $\Pi=-(-)^G$
states we have
\be
\Psi_\mu^{(G,-)}=
\pmatrix{ig_\mu^{(G,-;1)}(r)|G+1G+\frac{1}{2}GM\rangle \cr
-f_\mu^{(G,-;1)}(r) |GG+\frac{1}{2}GM\rangle \cr} +
\pmatrix{ig_\mu^{(G,-;2)}(r)|G-1G-\frac{1}{2}GM\rangle \cr
f_\mu^{(G,-;2)}(r) |GG-\frac{1}{2}GM\rangle \cr}.
\label{psineg}
\ee
Since the wave--functions are degenerate with respect to the
projection quantum number the $M$--dependence is not explicitly shown.
The radial functions $g_\mu$ and $f_\mu$ are obtained form diagonalizing
$h_{(0)}({\bf r})$. The boundary condition that the upper
components vanish at the edge of the spherical box is imposed,
\ie $g_\mu(D)=0$. Applying the transformation matrix
$u_{(0)}({\bf r})$ (\ref{u0}) onto the eigenfunctions
$\Psi_\mu^{(G,\pm)}$ alters the parity of these states.
Therefore we define new radial functions $\tilde g_\mu$ and
$\tilde f_\mu$ via
\be
u_0 \Psi_\mu^{(G,+)}=
\pmatrix{i\tilde g_\mu^{(G,+;1)}(r)|G+1G+\frac{1}{2}GM\rangle \cr
-\tilde f_\mu^{(G,+;1)}(r) |GG+\frac{1}{2}GM\rangle \cr} +
\pmatrix{i\tilde g_\mu^{(G,+;2)}(r)|G-1G-\frac{1}{2}GM\rangle \cr
\tilde f_\mu^{(G,+;2)}(r) |GG-\frac{1}{2}GM\rangle \cr},
\nonumber \\
\nonumber \\
u_0 \Psi_\mu^{(G,-)}=
\pmatrix{i\tilde g_\mu^{(G,-;1)}(r)|GG+\frac{1}{2}GM\rangle \cr
\tilde f_\mu^{(G,-;1)}(r) |G+1G+\frac{1}{2}GM\rangle \cr} +
\pmatrix{i\tilde g_\mu^{(G,-;2)}(r)|GG-\frac{1}{2}GM\rangle \cr
-\tilde f_\mu^{(G,-;2)}(r) |G-1G-\frac{1}{2}GM\rangle \cr}.
\label{psirot}
\ee
The action of the operators like ${\rm r}\cdot{\mbox{\boldmath $\tau$}}$
on the grand spin states $|ljGM\rangle$ is well known\cite{ka84} and
allows to relate the new radial functions to the originals, \eg
\be
\tilde g_\mu^{(G,+;1)}(r)=
-{\rm cos}\frac{\Theta}{2}f_\mu^{(G,+;1)}(r)
+\frac{1}{2G+1}{\rm sin}\frac{\Theta}{2}g_\mu^{(G,+;1)}(r)
-\frac{2\sqrt{G(G+1)}}{2G+1}
{\rm sin}\frac{\Theta}{2}g_\mu^{(G,+;2)}(r).
\nonumber
\ee
The strange quarks are not affected by the presence of the chiral
soliton. Thus the corresponding eigenstates are plane waves which
we decompose into spherical waves. We therefore write
\be
|1,\rho,j=l+\frac{1}{2},m\rangle _{\rm s}&=&{\cal N}_\rho^l
\pmatrix{i{\overline{w}}_{\rho l}^{\hskip 0.3em+}
j_l(k_{\rho l}r)|ljm\rangle \cr
{\overline{w}}_{\rho l}^{\hskip 0.3em-}
j_{l+1}(k_{\rho l}r)|l+1jm\rangle
\cr},
\nonumber \\* \nonumber \\*
|2,\rho,j=l-\frac{1}{2},m\rangle _{\rm s}&=&{\cal N}_\rho^l
\pmatrix{i{\overline{w}}_{\rho l}^{\hskip 0.3em+}
j_l(k_{\rho l}r)|ljm\rangle \cr
-{\overline{w}}_{\rho l}^{\hskip 0.3em-}
j_{l-1}(k_{\rho l}r)|l-1jm\rangle \cr}.
\label{stbasis}
\ee
The momenta $k_{\rho l}$ denote the roots
\be
j_l(k_{\rho l}D)=0
\label{discret}
\ee
which define the energyeigenvalues
${\overline{E}}_{\rho l}=\pm\sqrt{m_s^2+k_{\rho l}^2}$.
Furthermore the kinematical quantities
\be
{\overline{w}}_{\rho l}^{\hskip 0.3em+}&=&\sqrt{1+m_s/\hskip 0.3em
{\overline{E}}_{\rho l}},\quad {\overline{w}}_{\rho l}^{\hskip 0.3em-}=
{\rm sign}({\overline{E}}_{\rho l})
\sqrt{1-m_s/\hskip 0.3em {\overline{E}}_{\rho l}}.
\label{kinfac}
\ee
have been introduced. The factors ${\cal
N}_\rho^l=\big[D^{3/2}|j_{l+1}(k_{\rho l}D)|\big]^{-1}$ are determined
from normalization.

After having determined the eigenstates $|\mu\rangle$ of the
one--particle Hamiltonian $h_{(0)}$ we are able to compute
the matrix elements of $h_{(1)}$. The results are
\be
&&\langle \mu,G,M,+|\tilde h_{(1)}(\omega)| 1,\rho,l,m{\rangle_s}=
-\frac{m+m_s}{2}\delta_{G-1l}{\cal N}_\rho^{G-1}
\nonumber \\
&&\qquad \times
\int_0^D dr r^2
\Big[{\overline w}_{\rho G-1}^{\hskip 0.3em+}j_{G-1}(k_{\rho G-1}r)
\tilde g_\mu^{(G,+;2)}(r)+
{\overline w}_{\rho G-1}^{\hskip 0.3em-}j_{G}(k_{\rho G-1}r)
\tilde f_\mu^{(G,+;12}(r)\Big]
\nonumber \\
&&\qquad \times
\Big[\sqrt{\frac{G+M}{2G}}\delta_{Mm+1/2}a(r,\omega)
+\sqrt{\frac{G-M}{2G}}\delta_{Mm-1/2}b(r,\omega)\Big],
\label{h1p1} \\
\vspace{0.5cm}
&&\langle \mu,G,M,+|\tilde h_{(1)}(\omega)| 2,\rho,l,m{\rangle_s}=
-\frac{m+m_s}{2}\delta_{G+1l}{\cal N}_\rho^{G+1}
\nonumber \\
&&\qquad \times
\int_0^D dr r^2
\Big[{\overline w}_{\rho G+1}^{\hskip 0.3em+}
j_{G+1}(k_{\rho G+1}r) \tilde g_\mu^{(G,+;1)}(r)+
{\overline w}_{\rho G+1}^{\hskip 0.3em-}j_{G}(k_{\rho G+1}r)
\tilde f_\mu^{(G,+;1)}(r)\Big]
\nonumber \\
&&\qquad \times
\Big[-\sqrt{\frac{G-M+1}{2G+2}}\delta_{Mm+1/2}a(r,\omega)
+\sqrt{\frac{G+M+1}{2G+2}}\delta_{Mm-1/2}b(r,\omega)\Big],
\label{h1p2} \\
\vspace{0.5cm}
&&\langle \mu,G,M,-|\tilde h_{(1)}(\omega)| 1,\rho,l,m{\rangle_s}=
-\frac{m+m_s}{2}\delta_{Gl}{\cal N}_\rho^{G}
\nonumber \\
&&\qquad \times
\int_0^D dr r^2
\Big[{\overline w}_{\rho G}^{\hskip 0.3em+}j_{G}(k_{\rho G}r)
\tilde g_\mu^{(G,-;1)}(r)+
{\overline w}_{\rho G}^{\hskip 0.3em-}j_{G+1}(k_{\rho G}r)
\tilde f_\mu^{(G,-;1)}(r)\Big]
\nonumber \\
&&\qquad \times
\Big[-\sqrt{\frac{G-M+1}{2G+2}}\delta_{Mm+1/2}a(r,\omega)
+\sqrt{\frac{G+M+1}{2G+2}}\delta_{Mm-1/2}b(r,\omega)\Big],
\label{h1m1} \\
\vspace{0.5cm}
&&\langle \mu,G,M,-|\tilde h_{(1)}(\omega)| 2,\rho,l,m{\rangle_s}=
-\frac{m+m_s}{2}\delta_{Gl}{\cal N}_\rho^{G}
\nonumber \\
&&\qquad \times
\int_0^D dr r^2
\Big[{\overline w}_{\rho G}^{\hskip 0.3em+}j_{G}(k_{\rho G}r)
\tilde g_\mu^{(G,-;2)}(r)+
{\overline w}_{\rho G}^{\hskip 0.3em-}j_{G-1}(k_{\rho G}r)
\tilde f_\mu^{(G,-;2)}(r)\Big]
\nonumber \\
&&\qquad \times
\Big[\sqrt{\frac{G+M}{2G}}\delta_{Mm+1/2}a(r,\omega)
+\sqrt{\frac{G-M}{2G}}\delta_{Mm-1/2}b(r,\omega)\Big].
\label{h1m2}
\ee
These expressions are real and thus identical to their Hermitian
conjugate \hfil\break
${_s \langle} 1(2),\rho,l,m |\tilde h_{(1)}(-\omega)|
\mu, G, M, \pm\rangle$.

\medskip

\setcounter{equation}{0}
\renewcommand{\theequation}{B.\arabic{equation}}

\section*{Appendix B: Threshold behavior of the Dirac sea} \label{threshold}

\medskip

In this section we will derive the expression
$E_{th}=|\epsilon_{\rm val}|+m_s$ for the threshold which
stems from kaonic excitations of the quark states in the
Dirac sea. Here $\epsilon_{\rm val}$ denotes the smallest
(in magnitude) energy eigenvalue of $h_{(0)}$ (\ref{h0}) for a
state of vanishing strangeness.

A quark-- antiquark excitation is observed when the regulator
function of the fermion determinant acquires an imaginary part.
The regulator functions are of the form ({\it cf}. eqn.
(\ref{regfct}))
\be
\int_{1/\Lambda^2}^\infty ds s^{n/2}
R_i(s;\omega,\epsilon_\mu,\epsilon_\nu).
\label{regint}
\ee
These may be related to Feynman parameter integrals ($0\le x \le1$)
involving the incomplete error functions. The argument of which is
given by $\sqrt{X}$, with
\be
X = (1-x)\epsilon_\mu^2
+x\epsilon_\nu^2-x(1-x)\omega^2.
\label{argu}
\ee
The regulator functions thus acquire imaginary parts when the
inequality $X<0$ is satisfied (See \eg chapter 7 of
ref. \cite{ab65}.). $X$ possesses roots
\be
x_{1,2}=\frac{1}{2}
\left(1+\frac{\epsilon_\mu^2-\epsilon_\nu^2}{\omega^2}\right)
\pm\sqrt{-\frac{\epsilon_\mu^2}{\omega^2}
+\frac{1}{4}
\left(1+\frac{\epsilon_\mu^2-\epsilon_\nu^2}{\omega^2}\right)^2}.
\label{root}
\ee
The threshold, $E_{th}$, is determined by the condition
\be
E_{th}^2=\left(\epsilon_\mu\pm\epsilon_\nu\right)^2.
\label{realroot}
\ee
This has been obtained from the requirement that $X$ has exactly
one real root which is then given by
\be
x_1=x_2=\frac{\epsilon_\mu}{\epsilon_\mu\pm\epsilon_\nu}.
\label{x1x2}
\ee
The fact that the Feynman parameter integral only covers
the range $0\le x \le1$ further restricts the occurrence
of a threshold. There are two distinct cases
\begin{itemize}
\item
${\rm sign}\epsilon_\mu={\rm sign}\epsilon_\nu$. In this
case we need to take the upper sign yielding
\be
E_{th}^2=\left(\epsilon_\mu+\epsilon_\nu\right)^2
\qquad {\it i.e.}\qquad
E_{th}=|\epsilon_\mu|+|\epsilon_\nu|.
\label{uppersign}
\ee
\item
${\rm sign}\epsilon_\mu=-{\rm sign}\epsilon_\nu$. Hence the lower
sign is relevant
\be
E_{th}^2=\left(\epsilon_\mu-\epsilon_\nu\right)^2
\qquad {\it i.e.}\qquad
E_{th}=|\epsilon_\mu|+|\epsilon_\nu|.
\label{lowersign}
\ee
\end{itemize}
Of course, the physical threshold is the lowest possible value
for $E_{th}=|\epsilon_\mu|+|\epsilon_\nu|$.
For kaonic excitations one of these two eigenvalues corresponds to
a non--strange state while the other carries unit strangeness. The
lowest value for the former is given by $|\epsilon_\mu|=
|\epsilon_{\rm val}|$. For the latter it is just the strange
constituent quark mass $m_s$. Therefore we obtain for the
threshold originating form quark states in the Dirac sea
\be
E_{th}=|\epsilon_{\rm val}|+m_s
\label{thresproof}
\ee
as has been asserted. For $\epsilon_{\rm val}<0$ this is the lowest
threshold, whereas for $\epsilon_{\rm val}>0$ the singularity in
eqn. (\ref{valquad}) provides the lowest threshold as displayed in
Fig. 2.

\medskip

\vfil\eject

\end{document}